\documentclass[a4paper,11pt]{article}
\pdfoutput=1 

\usepackage{jheppub} 

\usepackage[T1]{fontenc} 

\usepackage{physics}

\usepackage{caption}

\usepackage{wrapfig}

\usepackage{mathrsfs}

\usepackage{slashed}

\usepackage[font=scriptsize]{caption} 

\usepackage{enumerate}

\usepackage{enumitem}

\usepackage{array}

\usepackage{makecell}

\usepackage{comment}

\usepackage{soul}

\usepackage{xcolor}     

\usepackage{mdframed} 

\usepackage{mathrsfs}

\usepackage[title]{appendix}

\usepackage{float}

\usepackage{subcaption}

\usepackage[normalem]{ulem}

\DeclareSymbolFont{yhlargesymbols}{OMX}{yhex}{m}{n} \DeclareMathAccent{\yhwidehat}{\mathord}{yhlargesymbols}{"62}

\newcommand*\df{\mathop{}\!\mathrm{d}} 

\title{\boldmath Geometry and Mechanics of Ribbon Gridshells }

\author[a]{Daniel Castro}
\author[b]{Joo-Won Hong}

\author[b]{\'Etienne Reyssat}
\author[b]{Jos\'e Bico}
\author[b]{Beno\^it Roman}
\author[a]{Hillel Aharoni}

\affiliation[a]{Department of Physics of Complex Systems, Weizmann Institute of Science, Rehovot 76100, Israel}
\affiliation[b]{Physique et M\'ecanique des Milieux H\'et\'erog\`enes (PMMH), ESPCI Paris, PSL University, CNRS, Sorbonne Universit\'e, Universit\'e Paris Diderot, Paris, France}

\emailAdd{hillel.aharoni@weizmann.ac.il}

\abstract{Mechanical metamaterials exhibit anomalous properties induced from non-trivial mesoscopic constituents. Inspired from architectural and industrial structures, we introduce arrangements of long, narrow ribbons intersecting at prescribed angles as a model thin-sheet metamaterial. These ribbon gridshells are shown to display highly non-linear behavior driven by the geometric constraints, nonetheless unlike most complex mechanical systems, we are able to explicitly write out the coarse-grained governing equations and to classify their solutions in terms of the intersection patterns of the ribbons. It is shown that the structure can assume arbitrary, tunable Gaussian curvature distributions, allowing us to formulate an inverse design problem, solvable under a suitably defined local condition. An analysis of the soft modes, confirmed by experiment and numerics, reveals rich mechanics which may exhibit both a rigid and an anomalously soft behaviors.}

\begin{document}

\maketitle
\flushbottom

\section{Introduction}\label{sec:intro}

Thin mechanical metamaterials have the ability to undergo large deformations driven by diverse actuation mechanisms. This functionality has fostered applications in fields like biomedicine \cite{Gao-16}, smart textiles \cite{Hu-12, Niu-25}, aeronautics \cite{Ajaj-16}, or architecture \cite{Ritter-06}, along with a modern understanding of multiple natural shape emerging processes \cite{Sharon-02,Armon-11,Sharon-07, Liang-09, Sharon-04, gladman-16}. Often in this framework, specific material properties are secondary to geometry and topology as the main determinants of the morphing \cite{mullerbook}. Prominent examples of geometric insight in mechanics include origami and kirigami (in \cite{tobasco-22, tobasco-23} and out of plane \cite{callens-18, Tani-24, sardas-25}), nematic \cite{nematic-review, Aharoni-14, Griniasty-19} and pneumatic \cite{Siefert-19, Siefert-20, Gao-20, Gao-23} elastomer sheets, hydrophilic gels \cite{Klein-07, Kim-12} or knitted and woven smart fabrics \cite{knittel-20, berin-26, Roy-23, Roy-25}.

On the other hand, in the recently emerged field of
architectural geometry \cite{Pottmann-15, book-shells} freeform design is understood and enhanced congenitally through differential geometry \cite{Pottman-07, Wallner-11, pottmann-16-book}. In particular, it has been shown beneficial for the fabrication and deployment of structures to align the supporting beams along the geodesic \cite{Mesnil-23, Pottmann-10, Weinand-06}, principal \cite{Pellis-18, Louikaides-14, Liu-23}, or asymptotic \cite{Schling-23-conf, Schling-18, Schling-18b, Block-24, Abaza-thesis} lines of the underlying surface. These assemblages, typically called \textit{gridshells}, are further termed Chebyshev gridshells or nets (Ch-nets) when all the beam segments have identical lengths. In this work, we only consider gridshells of this type. Besides their practicality, Ch-nets comprise a case study in extreme mechanics due to various exotic mechanical properties. Their complex shaping behavior when applied with various boundary conditions \cite{Baeck-18, Lefevre-15, saintjean-th}, and their mechanical response to strain in several scenarios \cite{Quaglierini-23} have been studied; they were proposed as well as models for stable reconfigurable surfaces \cite{Chen-21}.

In this work, we introduce the concept of \emph{ribbon gridshell} (RGS) as a simple metamaterial inspired by the Ch-nets used in architectural geometry and mechanics. RGSs are made of intersecting, non-stretchable ribbons arranged in a grid structure, with controlled intersection angles that serve as structural parameters (Fig.~\ref{fig:panel}a). Assembling a large number of long ribbons, we obtain a metamarial thin sheet with a unique parameter-dependent landscape of soft modes that allow it to conform into a variety of curved surfaces (Fig.~\ref{fig:panel}b). In this paper we study theoretically and experimentally their accessible geometries and soft modes. We further study the inverse design problem of choosing the intersection angles to allow shaping of the assembled surface into a desired shape. We find that tweaking the intersect assembly angles affects both the landscape of soft modes and the overall rigidity of the surface. We show that while most RGSs are almost entirely rigid, particular patterns of the intersection angles result in an anomalously soft highly-deformable surface.

\section{Surface geometry via continuum formulation   }\label{sec:formulation}

\begin{figure}[t!]
    \centering
 \hspace*{-1.3cm}
    \includegraphics[scale=0.4]{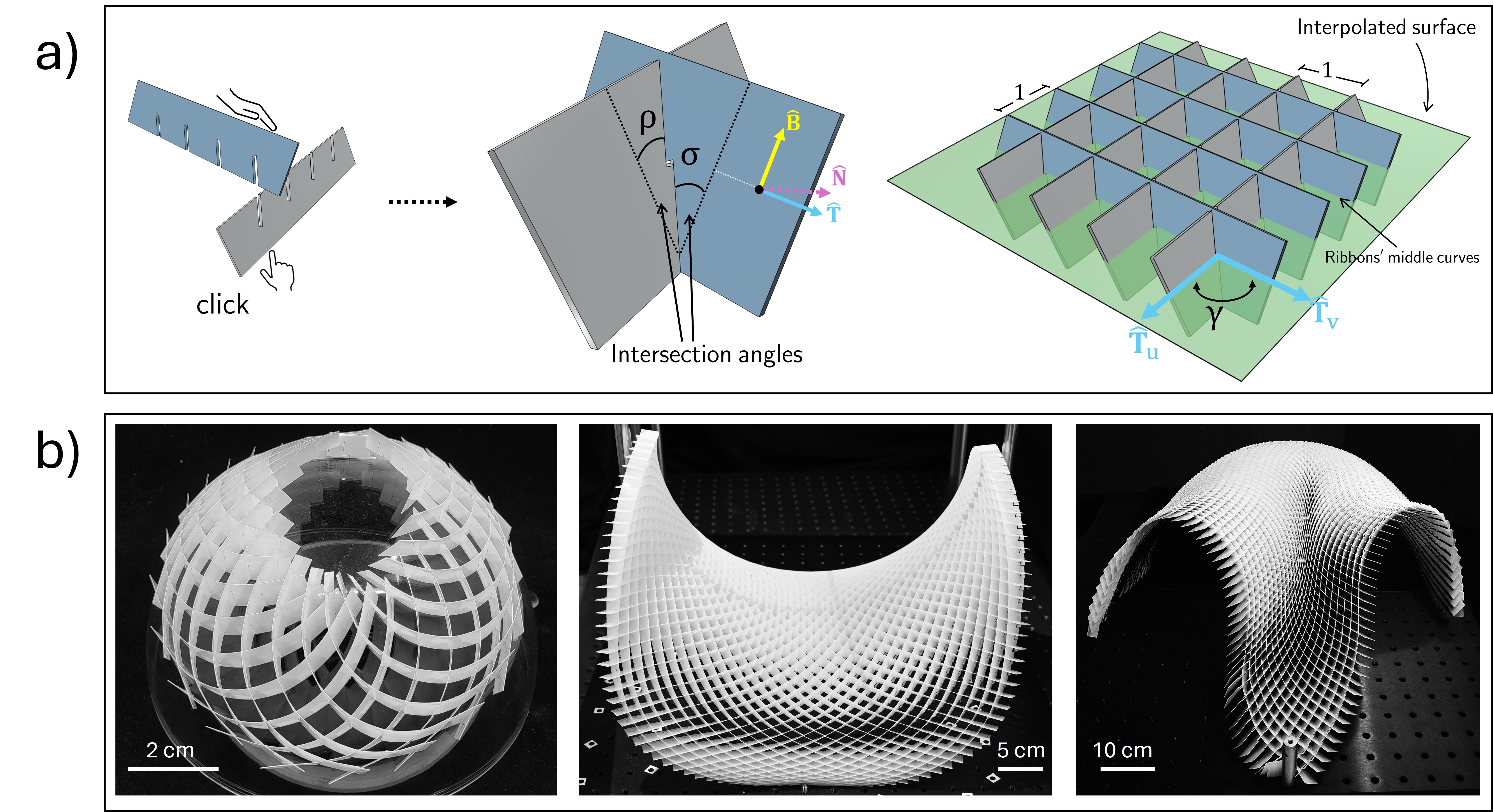}
    \caption{\textit{Ribbon gridshells} (RGSs). 
    a) Ribbons are laser-cut with a specified slitting pattern and interlocked on one another (left), the intersection line makes angles $\sigma$ and $\rho$ with the ribbons' binormals. Each ribbon lies on the plane spanned by its middle curve tangent and binormal vector (middle). The middle curves from each family form an angle $\gamma$ at the mutual contact points, the interpolation of the whole collection defining a continuous surface, shaded in green (right). b) These RGSs can conform to non-trivial shapes, with for instance, constant positive (left), constant negative (middle) or generally varying Gaussian curvature  (right).}
    \label{fig:panel}
\end{figure}

In this section we wish to derive the equations governing the shape of RGSs based on their local structural constraints, with tools from differential geometry. We consider gridshells made of ribbons ($\text{thickness} \ll \text{width} \ll \text{length}$) that intersect at pre-cut slits spaced equally along each ribbon's midline. The angle between a slit and the ribbons' width direction is denoted $\sigma$ (for one family of ribbons) and $\rho$ (for the transverse family). Each ribbon-ribbon intersection behaves as a rotating hinge, allowing the angle between the two ribbons to change freely.
Being thin elastic sheets, ribbons do not stretch longitudinally, nor do they curve within their own plane in order to avoid costly stretching deformations. However, they may twist and bend perpendicular to their plane \cite{grossman-16, efrati-11}. As an experimental realization of an RGS, we built several physical models made of hand-assembled laser-cut ribbons with different sequences of slit angles $\sigma, \rho$ (Fig. \ref{fig:panel}).

Zooming out, the assembly of many ribbons into a gridshell-like structure resembles another, larger thin sheet (Fig.~\ref{fig:panel}), whose geometry and mechanics are naturally determined by the properties of the constituent ribbons and their mutual interactions. Accordingly, in this limit we analyze the system with tools from thin sheet mechanics \cite{audoly-10-book, Efrati-09} and study common questions in this context: what is the deformation in response to an external forcing, which surface shapes can the sheet take, how these shapes are related to structural degrees of freedom, and how rigid or soft the system is with respect to different modes of deformation. In what follows, we derive the basic equations of RGSs, which will enable us to precisely formulate these questions. 

The mid-surface of an RGS is spanned by the midlines of the ribbons, and we conveniently use them as coordinate curves, denoting their arc-length parameters by $u$ and $v$. The configuration of the sheet is then given by $\mathbf{r}:{\Omega}\rightarrow\mathbb{R}^3$, with $\Omega\subset\mathbb{R}^2$ the $(u,v)-$domain of the assembled gridshell. Since we set the intersections between ribbons to be equidistant, and in the absence of longitudinal stretch, the choice of coordinates implies $|\mathbf{r}_u|=|\mathbf{r}_v|=1$, with subindex refering to partial derivative throughout the manuscript. Thus, the first fundamental form, or metric tensor, of the surface can be written as
\begin{align}
    \mathrm {I}=\df s ^2=\df u^2 +2\cos\gamma\: \df u \df v+\df v^2,
    \label{eq:metric}
\end{align}
where $\gamma=\gamma(u,v)$ is the local angle at which the midlines of ribbons from different families intersect. A surface patch with such distance constraints is called a \textit{Chebyshev net} \cite[pp.~100]{do-carmo} (Ch-net). The metric form \eqref{eq:metric} determines, by Gauss` \emph{Theorema Egregium}, the Gaussian curvature of the surface 
\begin{equation} 
K_{\mathrm{G}}=-\gamma_{uv}/\sin\gamma.
\label{eq:GaussCurvature}
\end{equation}
It otherwise contains no information about the embedding of the surface in $\mathbb{R}^3$. To study the shape of the embedded surface, we need the second fundamental form $\mathrm {I\!I}$, which chiefly depends on the ribbon properties and how they are joined at each pair of slits. The diagonal coefficients of $\mathrm {I\!I}$ -- the normal curvatures of the $u$ and $v$ coordinate lines -- are constrained by the ribbons' inability to bend in the direction of their width. By Meusnier’s formula \cite[p.~144]{shifrin,do-carmo}, they are proportional to the geodesic curvatures $\kappa_g^{(u)}=-\gamma_u$ and $\kappa_g^{(v)}=\gamma_v$, respectively. We can therefore write 
\begin{align}
    \mathrm {I\!I} = \gamma_u\:A(\gamma,\sigma,\rho)\,\df u^2 +2\tau\,\df u \df v+\gamma_v\:B(\gamma,\sigma,\rho)\,\df v^2.
    \label{eq:2-form}
\end{align}
The coefficients $A$ and $B$ are fully determined by the midline intersection angle $\gamma$ and by the local predetermined and fixed \emph{slit angles} $\sigma$ and $\rho$ (Fig.~\ref{fig:panel}, see appendix \ref{sec:app-formulas} for explicit formulae). The non-diagonal coefficient $\tau$ is the geodesic torsion along the $u$ and $v-$lines.

The only variable degrees of freedom in \eqref{eq:metric} and \eqref{eq:2-form} are thus $\gamma$ and $\tau$. The two are not entirely independent, since $\mathrm {I}$ and $\mathrm {I\!I}$ must obey the Gauss–Mainardi–Peterson–Codazzi (GMPC) constraints in order to describe an actual embedded surface \cite[p.~239]{do-carmo}. In terms of $\widetilde{\tau}\equiv\tau/\sin\gamma$ (namely $\tau$ per unit area), these nonlinear algebraic-differential equations take the form
\begin{subequations}
    \begin{align}
        \widetilde{\tau}_u&=\frac{\left( A\:\gamma_u\right)_v}{\sin\gamma}-B\:\frac{\gamma_u\gamma_v}{\sin^2\gamma} \label{eq:mpc1}\\
        \widetilde{\tau}_v &=\frac{\left( B\:\gamma_v\right)_u}{\sin\gamma}-A\:\frac{\gamma_v\gamma_u}{\sin^2\gamma}\label{eq:mpc2}\\
         \widetilde{\tau}^2&=\frac{\gamma_{uv}}{\sin\gamma}+AB\:\frac{\gamma_u\gamma_v}{\sin^2\gamma}.
         \label{eq:Gauss}
    \end{align}
    \label{eq:gmpc}
\end{subequations}

Strictly speaking, the study of the geometry of RGSs in the continuum framework amounts to the study of the system (\ref{eq:gmpc}). GMPC systems have been amply discussed in surface theory and thin sheet mechanics \cite{Poznyak-73, Poznyak-96, Efrati-09, Rozhdestvenskii}, however a general solution in this case may appear hard to reach, observing that the equation system is overdetermined (three equations for two unknowns) for generic values of the slit angles, and that the Codazzi–Mainardi equations (\ref{eq:mpc1})-(\ref{eq:mpc2}) contain second-order derivatives. Nonetheless, overdetermined systems often admit nontrivial solutions and, as detailed in Appendix \ref{sec:app-GMPC-analysis}, a complete classification is possible using PDE analysis methods. For any smooth $\sigma,\:\rho$, the system admits planar solutions with $\widetilde{\tau}=0$ and $\gamma$ constant. For some $\sigma,\:\rho$, there exist other non-trivial solutions (in addition to the planar ones), 
despite the overdeterminedness of the system (see case D in Appendix \ref{sec:app-GMPC-analysis}). If $\sigma,\:\rho$ are constant across the sheet, the overdeterminedness is lifted and we are left with a solvable system of either a single PDE or a set of two ODEs, depending on the particular values of $\sigma,\:\rho$. We elaborate on the simplest case within this family next. 

\paragraph{Asymptotic RGSs.}
If all the slits are perpendicular to the ribbon midline, we have $\sigma=\rho=0$ in eq.~\eqref{eq:2-form}. In this case, the ribbon normals (allowed bend directions) all lie in the tangent plane of the RGS surface, and thus ribbons cannot bend normal to it. As a result $A=B=0$, thus the ribbons follow the asymptotic lines of the surface, regardless of $\gamma$. We thus refer to this structure as an \emph{asymptotic RGS}.
From Eqs. \eqref{eq:gmpc} and \eqref{eq:GaussCurvature}, we deduce that $\widetilde{\tau}_u=\widetilde{\tau}_v=0$ and $-\widetilde{\tau}^2=K_\text{G}$, meaning that asymptotic RGSs exhibit constant non-positive Gaussian curvature $K_\text{G}\le 0$. The in-plane angle function, in turn, satisfies the sine-Gordon equation
\begin{align}
    \frac{\gamma_{uv}}{\sin\gamma}=\widetilde{\tau}^2=\text{const}\geq 0.
    \label{eq:sine-gordon}
\end{align}

\begin{figure}[h]
  \centering
  
  \includegraphics[width=1\textwidth]{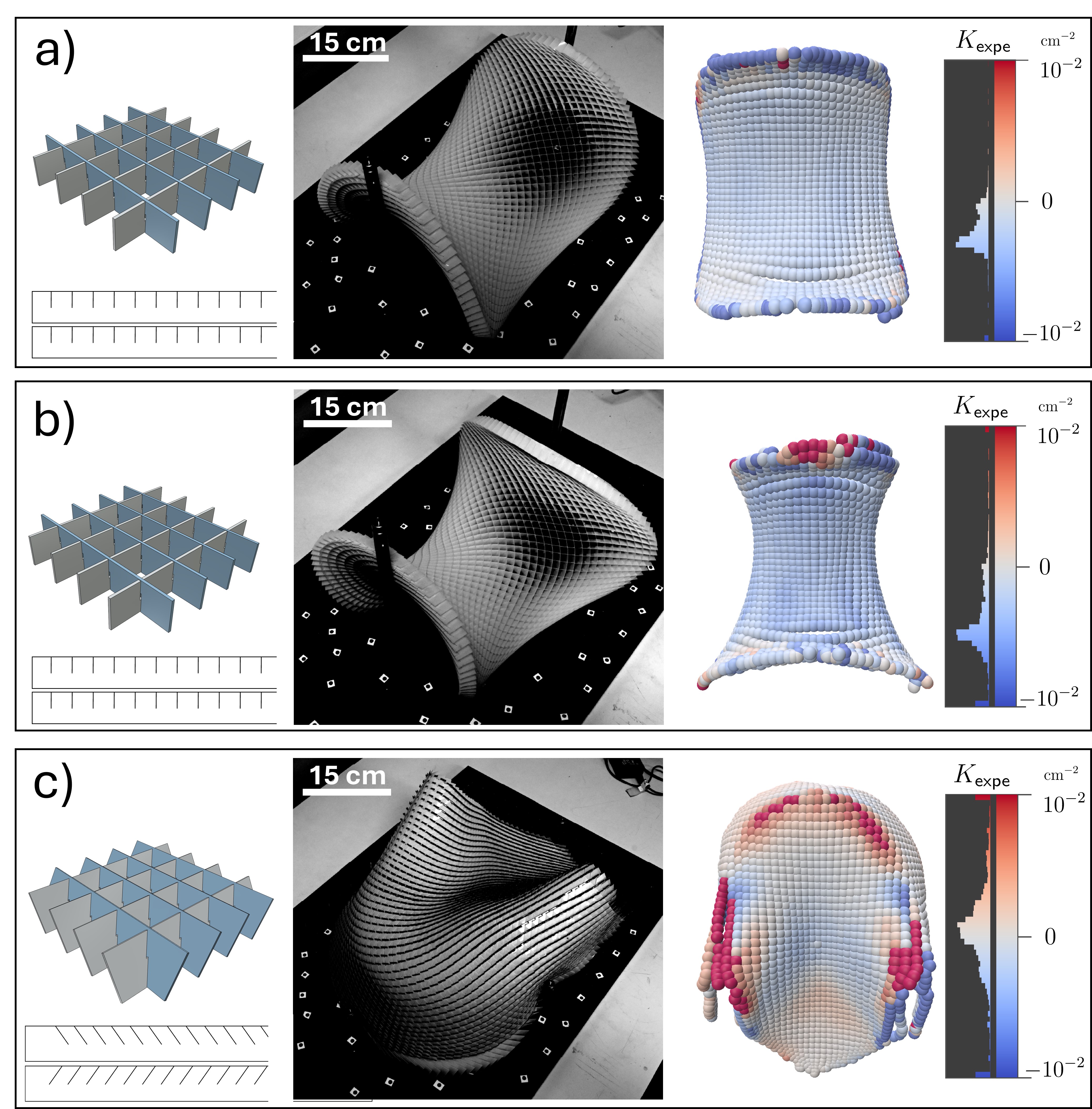}
  
\caption{Uniform slit angles RGSs. Scheme of ribbon patterns and assembled structure (left), a picture of a curved realization (middle), and the curvature-colored measured intersection positions (right). a) and b) Asymptotic RGS. In the absence of constraints the grid is floppy and can adopt different shapes depending on the way it is supported. Strikingly, any configuration exhibits uniform Gaussian curvature $K$ through the whole structure. However, the value of the constant depends on the configuration. The insets are colored by the Gaussian curvature $K$, which is accumulated around a single negative value. c) RGS with nonzero constant slit angles. $K$ varies across the surface, and the full distribution changes upon deformation.
}

  \label{fig:curvature_scans}
\end{figure}

We observe that although $\gamma(u,v)$ is a measure of the intrinsic geometry, it uniquely determines $\tau(u,v)$ and therefore the full shape of the surface and not just its Gaussian curvature. This intrinsic-extrinsic coupling is not a common feature in thin sheet systems; typically, the local structure and environmental coupling completely determine the intrinsic geometry (cf. \cite{Mostajeran-16, Griniasty-21}), and embedding selection is considered separately \cite{aharoni-18}. 
Experimental asymptotic RGSs are assembled by joining ribbons following an initial planar rectangular grid  ($\gamma=\pi/2$). We find that this planar state can easily be deformed into 3D non-nondevelopable shapes.
Fig.~\ref{fig:curvature_scans}a, b show two states of the same asymptotic RGS, with the measured Gaussian curvature for each of them. These experiments are consistent with the prediction that the field of Gaussian curvature is uniform, everywhere non-positive, but taking different values in a) and b). Note that even once the constant Gaussian curvature is determined, the intrinsic geometry of the sheet is highly flexible; the ribbons can display large geodesic curvatures, as seen in Fig.\ref{fig:blockers}a.

The GMPC set (\ref{eq:gmpc}) is thus reduced to the single equation (\ref{eq:sine-gordon}). It admits a unique solution given a Goursat condition \cite[p.~463]{courant1962methods}, namely the shape of a single curve from each of the two families. Changes in the this condition affect the overall shape, hinting at a highly non-local mechanics. In the case of vanishing curvature (fixing $\widetilde{\tau}=0$ in (\ref{eq:sine-gordon})) we get a planar sheet where the families of coordinate curves are congruent, i.e. pure translations of each other; this gives rise to novel applications in beam theory that are treated in a separate work \cite{Hong-26}, since our focus at present is on out-of-plane deformations.

When $\sigma,\:\rho$ are constants, with at least one of them nonzero, the Gauss' equation (\ref{eq:Gauss}) implies that the RGS surface will display non-constant curvature.  
For example, in Fig.~\ref{fig:curvature_scans}c an RGS with constant  nonzero $\sigma,\:\rho$ exhibits regions with both positive and negative Gaussian curvature.
Thus the GMPC system (\ref{eq:gmpc}) predicts quite dissimilar types of surfaces for different values of the input slit angles. 
All RGSs with uniform slit angles allow for a very wide family of configurations, as they depend on two one-variable functions (as a standard hyperbolic PDE, see Appendix \ref{sec:app-GMPC-analysis}). In contrast, only a finite number of constants determine the solutions in the case of non-constant $\sigma(u,v),\:\rho(u,v)$, as we show in sec.~\ref{sec:mechanics}. Is it possible to program the slit angle distribution to achieve a given target surface of arbitrary curvature?

\section{Inverse design}\label{sec:inverse}

In this section we address the inverse design problem---can \emph{any} surface be obtained by an RGS, and if so, how to compute the corresponding intersection angles $\sigma(u,v),\rho(u,v)$?

It is worth noting that any surface with uniform negative Gaussian curvature can be endowed with the metric (\ref{eq:metric}) satisfying (\ref{eq:sine-gordon}) and $A=B=0$ in (\ref{eq:2-form}) \cite{Gemmer-11,shankar2021}, and thus it can be obtained with asymptotic RGS ($\sigma=\rho=0$). Conversely, positive Gaussian curvature demands strictly non-vanishing $A$ and $B$ (equivalently, non-vanishing $\sigma$ and $\rho$). Thus, if realizable, it will result in a nontrivial non-asymptotic RGS. We thus pose a general inverse design problem, asking whether it is possible to construct an RGS that may take the shape of a desired surface. Given a target surface $S$, the inverse problem consists in finding fields $\sigma(u,v),\:\rho(u,v)$ such that fundamental forms (\ref{eq:metric})-(\ref{eq:2-form}) be defined on $S$. In general, however, for such functions not any $\gamma$ and $\widetilde{\tau}$ admit the required relations, but only a restricted compatible class. This is better illustrated as follows with the example of $S$ being a sphere.

If we seek $S$ to be a or part of sphere, the solutions to the GMPC system can be constructed analytically by setting 

\begin{align}
    A\left(\sigma,\rho\right)\cdot\gamma_u=B\left(\sigma,\rho\right)\cdot\gamma_v=\text{const}.
    \label{eq:sphere_angles}
\end{align}

From the system (\ref{eq:gmpc}) we then get

 \begin{align}
    \widetilde{\tau}=\text{const}\times\cot\gamma\:\:\:\:\:\:\:\text{and}\:\:\:\: \:\:\:\frac{\gamma_{uv}}{\sin\gamma}=-\text{const}^2.
    \label{eq:sine_gordon_flipped}
\end{align}

As before with (\ref{eq:sine-gordon}), we have $\widetilde{\tau}$ determined and a single, "flipped-sign" sine-Gordon PDE for $\gamma$. Eq.~(\ref{eq:sine_gordon_flipped}) implies that the first and second fundamental forms are linearly dependent, $ \mathrm {I\!I} =\text{const.}\times  \mathrm {I} $, thus describing either sphere of radius $R=1/\abs{\text{const}}$, or a plane if $\text{const}=0$  \cite[pp.~149]{do-carmo}. Nonetheless, a single Ch-net and thus a single RGS cannot cover the complete sphere \cite{Ghys, Samelson2018, Baeck-18}, and our construction is limited to a finite region. Contrary to the constant negative curvature case, the slit angles, obtained by inverting (\ref{eq:sphere_angles}), are now non-vanishing and depend explicitly on $\gamma$; this implies that for every particular realization a new set of angles must be used. In other words, here different $\gamma$ corresponds to different RGSs, as is the case for generic surfaces.

With arbitrary target surfaces, a closed form solution to the inverse problem is in general not possible. However, since we consider the middle curves of the ribbons as a Ch-net laid along the surface, we can start by constructing such a net (this is always possible by integrating a hyperbolic system, namely choosing an arbitrary line on the surface and expanding a Ch-net around it \cite{mason-17}). Then, as we show next, we can safely extrude these coordinate curves into ribbons when a suitable condition is met. In this way the inverse problem reduces mathematically to the existence of a restricted class of Ch-nets. In what follows we outline the general algorithm assuming such a condition (to be defined precisely below) is met.

Constructing a Ch-net on a target surface parametrized by $\mathbf{r}\left(\xi,\eta\right)$ amounts to finding a unit re-parametrization $u, v$; that is, $\mathbf{r}\left(\xi(u,v),\eta(u,v)\right)$ such that

\begin{align}
|\mathbf{r}_u|^2 =   E \:\xi_u^2+2F\: \xi_u \eta_u+  G \: \eta_u^2=1\:\:\:\:\:\:\:\text{and}\:\:\:\: \:\:\: |\mathbf{r}_v|^2&=    E \:\xi_v^2+2F\: \xi_v \eta_v+  G \: \eta_v^2=1,
\label{eq:norm}
\end{align}

with $E,\:F$ and $G$ the surface's first fundamental form coefficients. This can be solved in a variety of cases analytically \cite{ockendon_20, koendrink} and generally numerical methods are available to produce low distortion discrete nets \cite{Sageman-19, liu-20, liu-22, Oehri-24, Corman-25}. Once the Ch-net is at hand, we calculate the Frenet-Serret frame of each coordinate line, $\mathsf{FS}(u)=\{\mathbf{r}_u,\yhwidehat{\:\:\mathbf{r}_{uu}\:},\mathbf{r}_u\times \yhwidehat{\:\:\mathbf{r}_{uu}\:}\}\equiv\{\widehat{\mathbf{T}}^{(u)},\widehat{\mathbf{N}}^{(u)},\widehat{\mathbf{B}}^{(u)} \}$ and likewise for $v$, with the $\yhwidehat{\:\:\text{hat}\:\:}$ denoting normalization. The midcurve curvature vector lies along $\yhwidehat{\:\:\mathbf{r}_{uu}\:}$, and since the ribbon cannot bend in its plane, its cross section must be perpendicular to that direction. Therefore
the ribbons are defined as the surface spanned by $\widehat{\mathbf{T}}$ and $\widehat{\mathbf{B}}$ (more precisely, the $u$-ribbon at a fixed $v=v_*$ is given by $\mathsf{rib}(v_*)=\mathbf{r}(u,v_*)+h\:\widehat{\mathbf{B}}(u,v_*),\: h\in(-w/2,w/2)$, with $w$ the width, and similarly for $v$-ribbons). For a discrete net, usually obtained numerically when the surface does not have analytical parametrization, the equivalent calculation is described in the Appendix \ref{sec:app_explicit_inverse}. To obtain the slit angles we note that the ribbons intersect along the vector $\widehat{\mathbf{N}}^{(u)}\times \widehat{\mathbf{N}}^{(v)}$, so $\sigma$ and $\rho$ are (cf. Fig. \ref{fig:panel})

\begin{subequations}
    \begin{align}
\cos\sigma&=\widehat{\mathbf{B}}^{(u)}\cdot\yhwidehat{\:\:\widehat{\mathbf{N}}^{(u)}\times \widehat{\mathbf{N}}^{(v)}\:\:}\\
\cos\rho&=\widehat{\mathbf{B}}^{(v)}\cdot\yhwidehat{\:\:\widehat{\mathbf{N}}^{(u)}\times \widehat{\mathbf{N}}^{(v)}\:\:}.
    \end{align}
    \label{eq:angles_discr}
\end{subequations}

\begin{figure}[!t]
    \centering
\hspace*{-0.5cm}
    \includegraphics[scale=0.5]{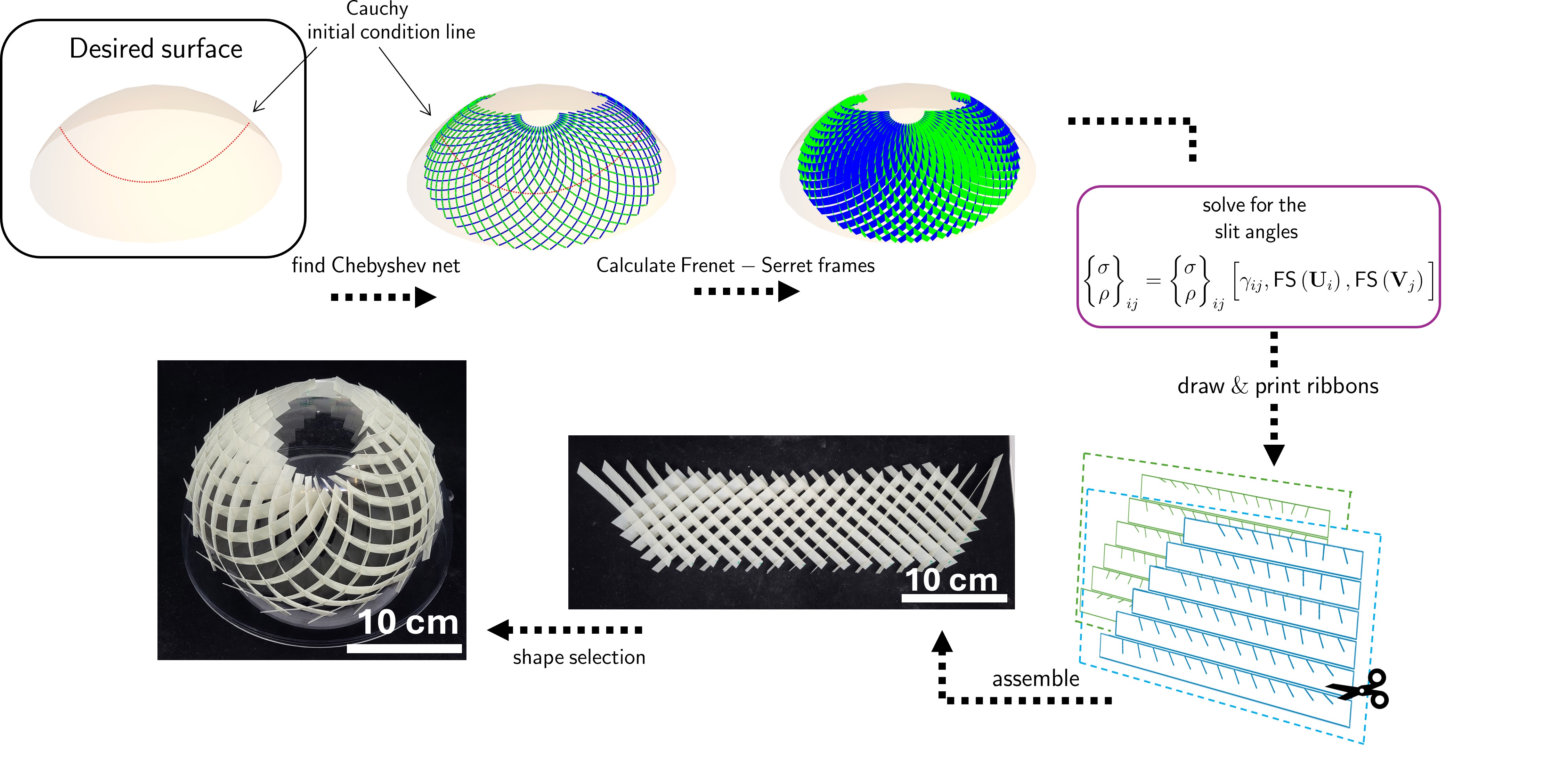}
    \caption{Making RGSs: A desired input surface must be provided with a Ch-net satisfying an initial/boundary condition along a line. The surface is thus reparametrized using as coordinates the net lines and from their Frenet-Serret frames we extract the distributions of slit angles $\sigma,\:\rho$. With the latter, the ribbons are laser cut on a foil of 
    Mylar (PET) and assembled  into the planar RGs. The final shape selection can be made by imposing adequate values of $\gamma$ or some equivalent fixing method such as inserting blockers. 
    }
    \label{fig:algorithm}
\end{figure}

We can thus summarize the protocol to solve the inverse problem for an input target surface (Fig.~\ref{fig:algorithm}). The protocol consists of four steps: (1) obtaining a Ch-net that covers the region of interest; (2) extruding the net lines into ribbons through the $\mathsf{FS}$ binormals; (3) calculating the slit angles using (\ref{eq:angles_discr}) and slitting a set of ribbons at those angles. (4) assembling the ribbons into an RGS, and deforming it with the appropriate $\gamma$ extracted from the first step, to correctly choose the target configuration.
In Fig.~\ref{fig:algorithm}, this protocol is demonstrated on a sphere, where the initial line for Ch-net construction is the $u=v$ line mapped to a latitude, along with constant values $\gamma(u,u)=1/2$ and $\gamma_n(u,u)=1/3$ along the line.

The algorithm described above may fall short of producing a practicable RGS in the following sense: if the midcurve geodesic curvature on the surface vanishes at any point, the normal vector for the ribbon aligns there with the surface normal, and the ribbon lies locally in the tangent plane of the surface. Therefore, a small variation will make the ribbon cross through the surface and consequently shift the slit angle by $180^\circ$, so our simple assembly method fails.
This problem can be mitigated, for instance by cutting slits on both sides of the ribbons and use some complex joint accordingly. In this case, our continuous treatment remains mostly unchanged. A practically easier resolution is to consider only patches of the Ch-net with non-vanishing geodesic curvatures, thus ensuring that $\sigma$ and $\rho$ change continuously between $-90^\circ$ and $90^\circ$. Recalling that the geodesic curvatures are precisely $-\gamma_u$ and $\gamma_v$, we must demand $\gamma_u\cdot\gamma_v\neq 0$ for an RGS-coverable region. In other words, the Ch-net curves must be constantly turning within the surface for them to be extruded into a feasible RGS. On an arbitrary surface, we can choose appropriate values of $\gamma$ and its derivatives as initial conditions for the Ch-net construction, guaranteeing satisfaction of the turning criterion in a finite region around the initial line. In Fig.~\ref{fig:algorithm}, the initial values $\gamma=1/2,\:\gamma_n=1/3$ where chosen by hand such that they produce an extended turning region. In conclusion, inverse design of generic target surfaces is locally possible, however global solutions may be hard to obtain and we have no rigorous results for the size of the solvable region.


\section{Rigidity and soft modes}\label{sec:mechanics}

In this section we aim to characterize the soft modes of RGS surfaces, and therefore their rigidity.
We approach this problem by trying to characterize the space of solutions to the system (\ref{eq:gmpc}) in the vicinity of a given solution. As we have shown, the system itself is overdetermined and thus may be expected to be rigid, nonetheless in some cases it is remarkably soft, such as the case of asymptotic RGSs, and more generally the case of constant $\sigma,\:\rho$. We next show that this degree of softness is not generic for all sets of slit angles, $\sigma(u,v),\:\rho(u,v)$, but rather stems from a hidden self-consistency implied by the uniform set (i.e. when $\sigma,\:\rho$ are constants).

To quantify this, we leave the realm of PDEs and switch back to a discrete analysis. Assuming a system of $N$-by-$N$ ribbons, there are $N^2$ ribbon-midline intersections, which we shall call vertices. Their positions in space amount to $N_\text{DoF}=3N^2$ degrees of freedom that fully capture the shape of the ribbon midlines (and thus the shape of the overall surface). In order for the grid to represent a Ch-net, each pair of neighboring vertices must be at unit distance from each other, imposing $2N(N-1)$ non-stretching constraints. In addition, every ribbon-ribbon intersection must intersect each of the two ribbons at the correct slit angle incised into it. Conveniently, at each non-boundary vertex, both ribbon planes are fully determined by the ribbon midlines; the ribbon's surface normal must coincide with the midline's Frenet-Serret normal $\widehat{\mathbf{N}}$ to avoid stretching of the ribbon surface. Thus, the $N_\text{DoF}$ position variables determine both ribbon slit angles at each non-boundary vertex, imposing $2(N-2)^2$ additional constraints, for a total of $N_\text{C}=2N(N-1)+2(N-2)^2=4N^2-10N+8$ constraints.

On the other hand, according to the Maxwell-Calladine theorem \cite{Calladine-78, Kane2013, mao_lubensky}, the number of soft modes in the system $N_\text{SM}$ is given by

\begin{equation}\label{eq:maxwell_counting}
	N_\text{SM}-N_\text{SS}=N_\text{DoF}-N_\text{C}=-N^2+10N-8,
\end{equation}
where $N_\text{SS}$ is the number of states of self-stress in the system. Since $N_\text{SM}$ is bound to be non-negative, we find that the former is of quadratic order, i.e $N_\text{SS}=\mathcal{O}\left(N^2\right)$, which is the discrete manifestation of the overdetermination of the system. 
Thus, since the slit angles are all arbitrary and independent, a generic RGS made of $2N$ ribbons cannot admit a nontrivial (non-constant $\gamma$) configuration.

One way to obtain a set of slit angles that can be realized into an RGS is to obtain them by inverse design, thus making sure that the $\mathcal{O}\left(N^2\right)$ states of self-stress are satisfied, resulting in $N_\text{SM}=\mathcal{O}\left(1\right)$ soft modes. Another way is to choose $\sigma, \rho$ such that the $2(N-2)^2$ slit angle constraints are dependent, thus resulting in states of self-stress that are trivially satisfied. This is the case of uniform angles, namely $\sigma(u,v)=\sigma_0$ and $\rho(u,v)=\rho_0$. In a separate manuscript (in preparation), we solve this case completely in a continuum setting. Similar to Eq.~(\ref{eq:sine-gordon}), the geometric inter-dependencies reduce the system to a single scalar hyperbolic PDE. Such a system is uniquely solvable if augmented with a scalar initial condition along two characteristic curves (a Goursat problem), or a two-scalar initial condition along a single non-characteristic curve (a Cauchy problem). In the discrete setting, this translates to $2N+\mathcal{O}(1)$ soft modes (in the absence of additional constraints or nontrivial states of self-stress).

\paragraph{Experiment.}

\begin{figure}[t!]

  \centering
\includegraphics[width=1\textwidth]{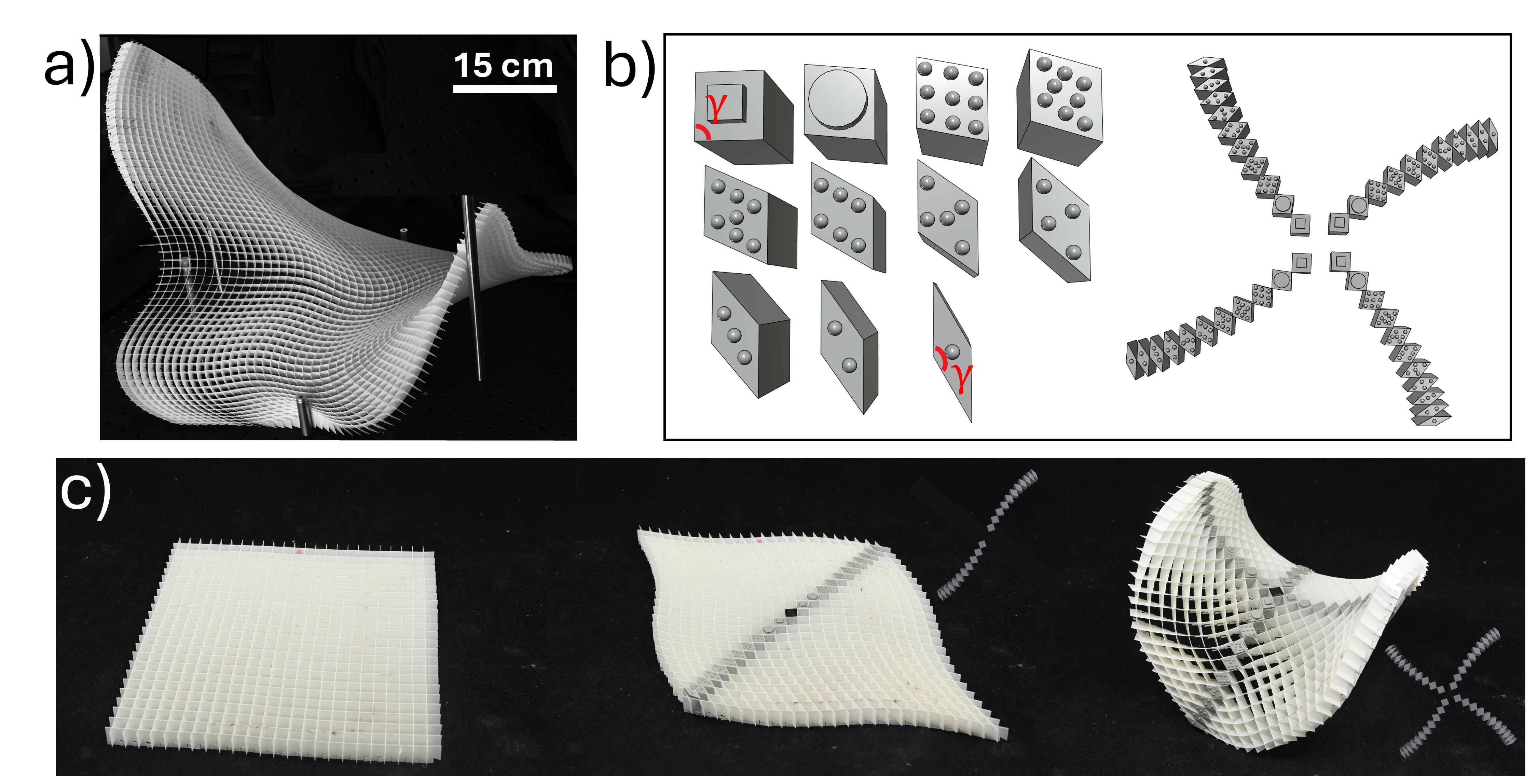}
  
  \caption{a) A large asymptotic RGS exhibiting a state of constant negative Gaussian-curvature and a complex distribution of ribbon geodesic curvatures. The shape is only selected by gravity and external boundary constraints, akin to the common dogma in architecture. b) Small 3d printed diamond faced prisms can be set as blockers to select a shape; each blocker adds a new constraint in Eq.~(\ref{eq:maxwell_counting}). Blockers of different $\gamma$-angles can be used, and are denoted by different bulge-coded labels. c) If arranged along both diagonals, no additional constraints are needed, producing a fully rigid asymptotic RGS, in line with the softness analysis discussed in this section. An insufficient number of blockers (as in the middle picture) preserves some softness.}

  \label{fig:blockers}
\end{figure}

An embedding selection mechanism, convenient as well to corroborate experimentally the counting of soft modes, consists in imposing additional constraints in the form of blockers that fix the local angle $\gamma$ at a single point (Fig.  \ref{fig:blockers}). These can be designed for generic slit angles, however for simplicity we considered only asymptotic RGSs where the blockers are diamond-faced prisms. Set along the central ribbon of each family (thus mirroring Goursat type initial conditions), these constraints remove the remaining $2N$ soft modes consistently, so that $N_\text{SS}=0$, yielding a completely rigid 3d structure ($N_\text{SM}=0$). 

\begin{figure}[t!]

  \centering
\hspace*{-1.4cm}
\includegraphics[width=1\textwidth]{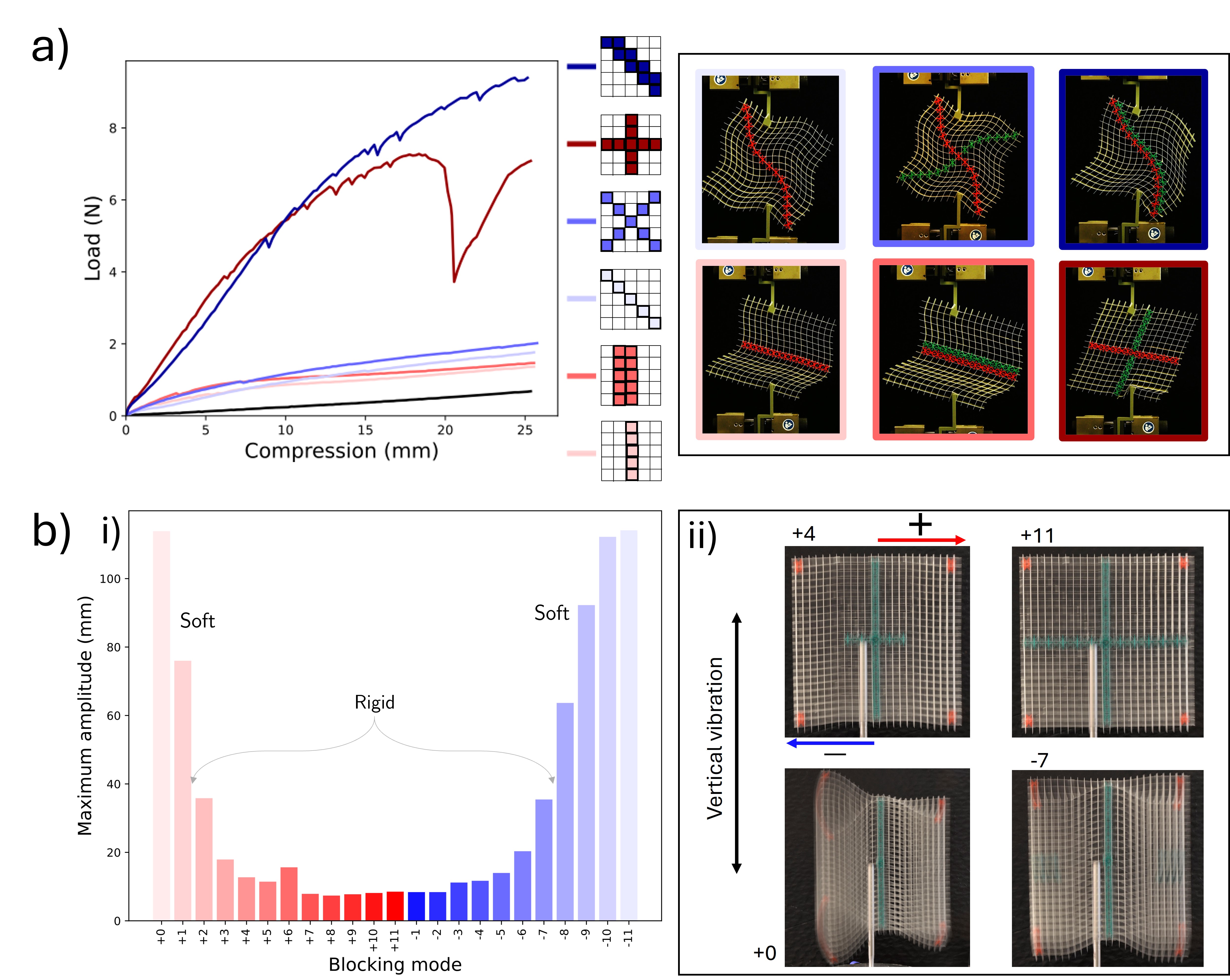}

\caption{a) Compression test on grids of dimension $15 \times 15$ cm with constant spacing of $1$ cm for various distributions of blockers, as schematized and pictured on the right with colors corresponding to the plot. The loading is performed with the grid oriented at 22.5$^\circ$ to avoid compression along diagonals or lines. A small load-to-displacement ratio indicates softness, as it is observed with the unconstrained configuration (black line). Some arrangements of constraints also appear considerably soft (light lines), while 
two in particular display much higher force-to-displacement ratio, i.e. are rigid (dark lines). b) Vertical vibration test with one full alignment of blockers already placed along the vertical (for the same grid as in a)). Blockers are added from the center to the boundary (+ in red). Then, the blockers are removed from the center to the boundary (- in blue). Pictures of the grids show from left to right, top to bottom: $+4$, $+11$, $0$ (or $-11$) and $-7$. We observe the marked increase of softness when less blockers are placed in the grid.
}

\label{fig:vibrations}
\end{figure}

In order to measure the softness of RGSs we perform two sets of measurements. In the first, asymptotic RGSs are compressed incrementally while the exerted force is measured (Fig. \ref{fig:vibrations}-a). We place different arrangements of square-faced blockers and measure in each case the elastic response of the constrained systems. We find that, indeed, with less than $2N$ blockers, the system maintains its floppiness (pink and lavender-blue curves), which is indicated by small force to displacement ratios. On the contrary, $2N+\mathcal{O}(1)$ blockers rigidify the system when arranged as a Goursat or Cauchy initial condition (dark blue and red curves, respectively), displaying a much steeper force-displacement curve. If arranged differently, some of the additional constraints may create states of self-stress rather than eliminating soft modes (light blue and salmon curves). In these cases, $N_\text{SM}-N_\text{SS}=0$; however, $N_\text{SM}>0$  and therefore the system is floppy.

In the second experiment RGSs are connected to a harmonically oscillating rod along a next-to-middle ribbon and the amplitude of the vibrations is recorded, with different numbers of blockers being placed horizontally, away and toward the center, as shown in Fig. \ref{fig:vibrations}-b. The notations $+$ and $-$ indicate that a blocker is placed or removed from a cell. Naturally when there is no blocker ($+0$ and $-11$) the amplitude is at its maximum, and it goes decreasing with the number of adjoined blockers, showing how the softness of the structure is dramatically reduced as the limit $N_\text{SM}=0$ of no soft-mode is approached.

\paragraph{Numerics.} 

The experiments described above were carried out exciting solely the planar modes, however the counting following Eq.(\ref{eq:maxwell_counting}) applies as well to out-of-plane deformations. Moreover, since a direct experimental testing of those buckling modes is more challenging, we perform instead a numerical verification using simple geometric rigidity analysis. On a given curved surface we construct a Ch-net with $N^2$ vertices, which we call base net, and consider linear perturbations around this state. As the base net, every perturbed net has associated a set of slit angles at each vertex, $\sigma_{ij}$ and $\rho_{ij}$, which are functions of the vertex coordinates $\mathbf{r}_{ij}$.
When perturbed, we penalize quadratically deviations from the constraints, which are the base net angles and edge lengths:

\begin{align*}
    \mathsf{Energy}=\sum_{\expval{(ij),(kl)}}\left(|\mathbf{r}_{ij}-\mathbf{r}_{kl}|-L \right)^2+\sum_{[i,j]}\left[\left( \sigma_{ij}-\sigma_{ij}^\text{B}\right)^2+\left( \rho_{ij}-\rho_{ij}^\text{B}\right)^2\right]
\end{align*}
with $\expval{(ij),(kl)}$ and $[i,j]$ denoting respectively edges and non-boundary vertices, and the subscript $^\text{B}$ the fixed base values. To characterize the spectrum of soft modes we extract the eigenvalues of the Hessian matrix of $\mathsf{Energy}$, which around equilibrium is simply $\mathsf{H}=J^\mathsf{T}J$, with $J$ the Jacobian of the constraints.

\begin{figure}[t!]

  \centering
\hspace*{-1.0cm}
    \includegraphics[scale=0.2]{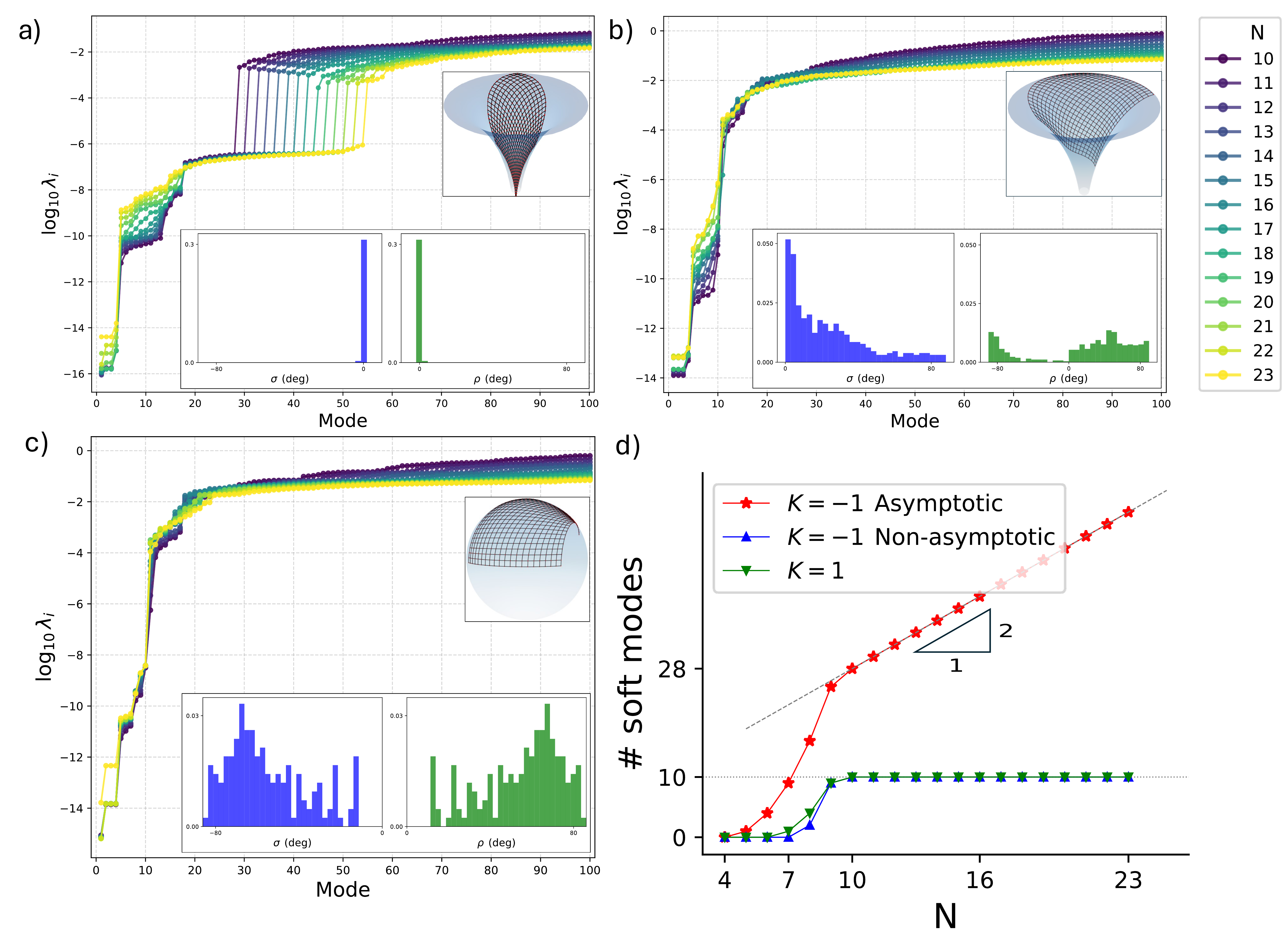}
    \caption{Distribution of the singular values $\lambda_i$ of the Jacobian $J$. (Since the Hessian is $H=J^\mathsf{T}J$, its eigenvalues are $\lambda_i^2$.) The smaller values accumulate below a threshold of around $10^{-6}$ and correspond to soft modes, after which a large gap separates the finite energy modes. If the slit angles are uniform, as in a pseudospherical surface with asymptotic Ch-net (a), the number of soft modes increases linearly with $N$ at a slope of exactly 2. (b) The same surface might be covered with a non-asymptotic Ch-net, showing instead an $N-$independent distribution. When the slit angles change appreciably from vertex to vertex (see the inset histograms for $\sigma$ and $\rho$) the number of soft modes converges to a constant value at $N=10$, as in (b) or in the case of a sphere (c). (d) In each case, the number of soft modes is counted as a function of $N$. The profile of the distribution depends primarily on $\sigma,\:\rho$, not the surface itself.}
    \label{fig:spectrum}
    
\end{figure}

We observe (Fig. \ref{fig:spectrum}) two characteristic patterns, exemplified with base surfaces of a constant-curvature saddle and a sphere. When the slit angles are uniform (the case of the saddle with asymptotic Ch-net Fig. \ref{fig:spectrum}.a), the distribution of eigenvalues reveals a large gap which separates low stiffness modes (soft modes) from high stiffness ones. The number of soft modes (the position of the gap) increases linearly with  $N$. In contrast, for non-constant slit angles (both the non-asymptotic Ch-net on the saddle \ref{fig:spectrum}.b and a Ch-net on the sphere  \ref{fig:spectrum}.c) the distribution is almost identical for every $N$, with the gap occurring at the 10th mode. The soft modes correspond to eigenvectors of the Hessian with eigenvalues under a sufficiently low threshold ($10^{-12}$ in Fig. \ref{fig:spectrum}).

This behavior corroborates the continuum limit and the analysis following Eq. (\ref{eq:maxwell_counting}). If $\sigma,\:\rho$ are uniform, $\gamma$ and $\widetilde{\tau}$ are determined by two initial-value functions of a hyperbolic problem (equivalent to $2N+\mathcal{O}(1)$ discrete independent constraints), thus the number of soft modes increases like $2N$, as in Fig. \ref{fig:spectrum}.a. On the other hand, if $\sigma,\:\rho$ vary, the solution for $\gamma$ and $\widetilde{\tau}$ depends on constants, not initial-value functions, so the number of soft modes is likewise $\mathcal{O}(1)$, as in  Fig. \ref{fig:spectrum}.b-c.

\section{Discussion}

Ribbon Gridshells (RGSs) were studied in this paper not because they are of crucial importance in any particular way. Rather, RGSs represent a class of two-dimensional metamaterials whose intrinsic local structure consists of a single-DOF mechanism, that is strongly coupled to both their in-plane geometry and out-of-plane curvatures. This results in highly-non-local soft modes that heavily restrict the family of shapes and deformations accessible for these structures. Tweaking the local mechanism, in our case via the ribbon slit angles, allows inverse-design of soft modes and shapes. Nonetheless, geometric relations render some of these structures almost rigid while others, such as our asymptotic RGSs, remain significantly more compliant.

Previous works featuring curvature generation in thin sheet structures often address systems of constant curvatures (e.g. periodic shells \cite{Nassar-17, nassar-24}). On the other hand, systems with varying local curvature generation capabilities (e.g. \cite{Roy-23,knittel-20}) typically necessitate ad hoc embedding selection mechanisms \cite{aharoni-18}. Instead, RGSs pose a model for systems in which intrinsic and extrinsic degrees of freedom are coupled and governed by a highly non-local, though controllable, mechanics. Our inverse design approach is shape-oriented, namely it inherently addresses both the two-dimensional geometry and its space embedding simultaneously.

As we showed in sec. \ref{sec:inverse}, practicable RGS inverse design for simple slit ribbons requires constructing a "constantly-turning" Ch-net, namely one where the geodesic curvatures of the curves are non-vanishing at every point. On a given surface, infinitely many Ch-nets may be integrated by changing the initial-value conditions for their generation, even globally in some cases \cite{mason-17}. However, there is no known result bounding the region of feasibility for the always-turning condition. However, variational approaches like the ones in \cite{Sageman-19, liu-20, liu-22, Oehri-24, Corman-25} may be adjusted to incorporate the condition by adding a suitably defined energy term to the Ch-net energy functional \cite[sec.~6.6]{Corman-25}. A refined numerical inverse problem protocol incorporating this adjustment is in preparation.

In sec. \ref{sec:mechanics} we observed that for slit angles obtained by inverse design, thus compatible with an actual shape, there is generically a small $\mathcal{O}(1)$ number of soft modes, independent of the system size. These modes can be eliminated to make a rigid structure by setting just a few additional constraints. In this sense, RGSs are a model system for a surface design principle, where almost rigid surfaces are deformable only along very few devised paths (cf. for instance \cite{liu-shokef,Chen-18}). In contrast, asymptotic RGSs exhibit a number of soft modes that scales linearly with the system size. The full manifold of accessible shapes can be foliated into Gaussian-curvature-indexed submanifolds, each of which corresponds to the full space of solutions to the sine-Gordon equation (\ref{eq:sine-gordon}). We propose the RGS model to advance the understanding of soft modes and surface deformation paths in thin sheet systems. In particular, we propose an approach based on the inverse design of soft modes, instead of classical single-shape inverse design approaches which lie at the forefront of the field of shape-shifting matter.

\section{Acknowledgments}

We thank Raz Kupferman and Roee Leder for useful discussions on the mathematical formulation in sec. \ref{sec:formulation}. This research was
supported by the Weizmann-CNRS collaboration program DESMOD.

\section{Statements}

The authors declare no conflict of interest.\\

Data and code are available upon requests.

\appendix

\section{Explicit formulas}\label{sec:app-formulas}

\subsection{Derivation of the second fundamental form (\ref{eq:2-form})}

In general the diagonal entries of the second fundamental form can be conveniently written as $L=\kappa^{(u)}\cos\phi^{(u)}$ and $N=\kappa^{(v)}\cos\phi^{(v)}$, $\phi$ being the angle between the curve normal and the surface normal and $\kappa$ the curvature of the respective coordinate lines. The latter can be eliminated in favor of the geodesic curvature, which depends only on the metric, observing that $\kappa_g=\kappa\:\sin\phi$, so $L=\kappa^{(u)}_g\cot\phi^{(u)},\:N=\kappa^{(v)}_g\cot\phi^{(v)}$. Plugging in the metric we obtain $\kappa_g^{(u)}=-\gamma_u$ and $\kappa_g^{(v)}=\gamma_v$, while the coefficients can be written as

\begin{subequations}
    \begin{align}
   \cot\phi^{(u)} &=\frac{\widehat{\mathbf{N}}_s\cdot\widehat{\mathbf{N}}^{(u)}}{\sqrt{1-(\widehat{\mathbf{N}}_s\cdot\widehat{\mathbf{N}}^{(u)} )^2}}\equiv -A \\
      \cot\phi^{(v)} &=\frac{\widehat{\mathbf{N}}_s\cdot\widehat{\mathbf{N}}^{(v)}}{\sqrt{1-(\widehat{\mathbf{N}}_s\cdot\widehat{\mathbf{N}}^{(v)} )^2}}\equiv B. 
\end{align}
\end{subequations}

where $\widehat{\mathbf{N}}_s=\widehat{\mathbf{T}}^{(u)}\times\widehat{\mathbf{T}}^{(v)}\Big/\sin\gamma$ is the surface normal. These dot products can be further expressed in terms of the tangents and binormals which span the ribbons; observing that the intersection of the slits is along the vector

\begin{align*}
    \sin\sigma \:\widehat{\mathbf{T}}^{(u)}+\cos\sigma \:\widehat{\mathbf{B}}^{(u)}= \sin\rho \:\widehat{\mathbf{T}}^{(v)}+\cos\rho \:\widehat{\mathbf{B}}^{(v)}
\end{align*}

we have

\begin{subequations}
    \begin{align} -\widehat{\mathbf{N}}_s\cdot\widehat{\mathbf{N}}^{(u)}&=\frac{\widehat{\mathbf{T}}^{(v)}\cdot\widehat{\mathbf{B}}^{(u)}}{\sin\gamma}=\frac{\sin\rho-\sin\sigma\cos\gamma}{\sin\gamma\cos\sigma}\\   \widehat{\mathbf{N}}_s\cdot\widehat{\mathbf{N}}^{(v)}&=\frac{\widehat{\mathbf{T}}^{(u)}\cdot\widehat{\mathbf{B}}^{(v)}}{\sin\gamma}=\frac{\sin\sigma-\sin\rho\cos\gamma}{\sin\gamma\cos\rho},
\end{align}\label{eq:app:normal_prods}
\end{subequations}

therefore $ \mathrm {I\!I} = \gamma_u\:A(\gamma,\sigma,\rho)\,\df u^2 +2\tau\,\df u \df v+\gamma_v\:B(\gamma,\sigma,\rho)\,\df v^2$ with

\begin{subequations}
    \begin{align}
A(\gamma,\sigma,\rho)&=\frac{\sin\rho-\sin\sigma\cos\gamma }{\sqrt{\cos^2\sigma\sin^2\gamma-\left(\sin\rho-\sin\sigma\cos\gamma \right)^2}} \\
B(\gamma,\sigma,\rho)&=\frac{\sin\sigma-\sin\rho\cos\gamma }{\sqrt{\cos^2\rho\sin^2\gamma-\left(\sin\sigma-\sin\rho\cos\gamma \right)^2}}.
\end{align}\label{eq:app:A_B}
\end{subequations}

Note that $A(\gamma,\sigma,\rho)=B(\gamma,\rho,\sigma)$, reflecting the symmetry of the system under $u,\sigma\longleftrightarrow v,\rho$. On the other hand, the off-diagonal entry $\tau$ cannot be written in terms of the metric and its derivatives, so effectively it is the only independent variable in $ \mathrm {I\!I}$.

\subsection{Inverse problem}\label{sec:app_explicit_inverse}

In this sub-section we show the formulae for the slit angles as described in the sec. \ref{sec:inverse}, first for the case of a target surface that has analytic parametrization and also when the Ch-net is given as an ordered point cloud.

 If a parametrized surface $\mathbf{r}\left(\xi,\eta\right)$ is given a Ch-net, i.e. a $u,v$ reparametrization satisfying (\ref{eq:norm}), the angle function can be extracted with 

\begin{align}
    \cos\gamma(u,v)=E\:\xi_u \xi_v+F\left(\xi_u \eta_v+\xi_v\eta_u \right)+G\:\eta_u \eta_v
    \label{eq:app:cos_g}
\end{align}

and the surface normal as

\begin{align}
    \widehat{\mathbf{N}}_s=\frac{\xi_u\eta_v-\xi_v\eta_u}{\sin\gamma}\left(\frac{\partial \mathbf{r}}{\partial \xi}\times \frac{\partial \mathbf{r}}{\partial \eta}\right).
\end{align}

From these equalities we obtain the diagonal second fundamental form coefficients:

\begin{subequations}
   \begin{align}
        \sin\gamma\:  \widetilde{L}=&  \widehat{\mathbf{N}}_s\cdot \frac{\partial^2 \mathbf{r}}{\partial u^2}=  \left( \xi_u\right)^2 \:L^{(\xi,\eta)} +2\xi_u\eta_u  \: M^{(\xi,\eta)}+\left( \eta_u\right)^2 \:N^{(\xi,\eta)}=\gamma_u\:A(\gamma,\sigma,\rho) \\
               \sin\gamma  \:\widetilde{N} =&  \widehat{\mathbf{N}}_s\cdot \frac{\partial^2 \mathbf{r}}{\partial v^2}= \left( \xi_v\right)^2 \:L^{(\xi,\eta)} +2\xi_v\eta_v  \: M^{(\xi,\eta)}+\left( \eta_v\right)^2 \:N^{(\xi,\eta)}=\gamma_v\:B(\gamma,\sigma,\rho),
   \end{align}
   \label{eq:indu_lm}
\end{subequations}

where the index $^{(\xi,\eta)}$ refers to the quantities in the natural coordinates $\xi$ and $\eta$, and $\widetilde{\:\Box\:}=\Box/\sin\gamma$. The non-diagonal component reads

\begin{align}
       \sin\gamma\:  \widetilde{\tau}=&  \widehat{\mathbf{N}}_s\cdot \frac{\partial^2 \mathbf{r}}{\partial u \partial v}=\xi_u \xi_v\:L^{(\xi,\eta)} +\left(\xi_u \eta_v+\xi_v \eta_u \right)M^{(\xi,\eta)} +\eta_u \eta_v\:N^{(\xi,\eta)}.
\end{align}

The left hand side of (\ref{eq:indu_lm}) is in turn written in terms of $\sigma,\:\rho$, recalling (\ref{eq:app:A_B}), and $\gamma$, the latter already determined by (\ref{eq:app:cos_g}). Thus replacing the explicit expressions and solving we arrive at

\begin{subequations}
    \begin{align}
    \sin\sigma &=\frac{\frac{ \widetilde{N}}{\gamma_v}-\frac{ \widetilde{L}}{\gamma_u}\cos\gamma}{\sqrt{1+\left(\frac{ \widetilde{L}}{\gamma_u}\right)^2+\left(\frac{ \widetilde{N}}{\gamma_v}\right)^2-2\frac{ \widetilde{L}}{\gamma_u}\frac{ \widetilde{N}}{\gamma_v}\cos\gamma}}\\
    \sin\rho &=\frac{\frac{ \widetilde{L}}{\gamma_u}-\frac{ \widetilde{N}}{\gamma_v}\cos\gamma}{\sqrt{1+\left(\frac{ \widetilde{L}}{\gamma_u}\right)^2+\left(\frac{ \widetilde{N}}{\gamma_v}\right)^2-2\frac{ \widetilde{L}}{\gamma_u}\frac{ \widetilde{N}}{\gamma_v}\cos\gamma}}.
\end{align} \label{eq:sol_angles_cont}
\end{subequations}

In this way the slit angles are written in terms of the natural coordinates through the second fundamental diagonal coefficients and their derivatives.\\

\begin{table}
\centering
\renewcommand{\arraystretch}{1.8}
\begin{tabular}{ll}
\underline{\text{ Discrete Frenet-Serret $\mathsf{FS}(\mathbf{U}^{(i)}_j)$ frame:}} &  \\ 
 For each point $i$ take the $j$-derivative which is defined on edges: & $\df\mathbf{U}^{(i)}_j=\frac{\mathbf{U}^{(i)}_{j+1}-\mathbf{U}^{(i)}_j}{\epsilon}$ \\
The average of $\df\mathbf{U}^{(i)}_j$ is proportional to the tangent at the point: & $\mathbf{T}^{(i)}_{j+1}=\frac{\df\mathbf{U}^{(i)}_j+\df\mathbf{U}^{(i)}_{j+1}}{2}$ \\
 The second $j$-derivatives of $\mathbf{U}^{(i)}_j$ are proportional to the normals: & $\mathbf{N}^{(i)}_{j+1}=\df\df\mathbf{U}^{(i)}_j$ \\
 The binormals are the cross product of tangents with normals: & $\mathbf{B}^{(i)}_{j}=\mathbf{T}^{(i)}_{j}\times \mathbf{N}^{(i)}_{j}$ \\
 At each point the triad must be orthonormalized: & \makecell{$\begin{aligned}
 \mathsf{FS}(\mathbf{U}^{(i)}_j)=&\mathscr{ON}\left[\mathbf{T}^{(i)}_{j},\mathbf{N}^{(i)}_{j},\mathbf{B}^{(i)}_{j} \right]\\
\equiv&\left\{\widehat{\mathbf{T}}^{(i)}_{j},\widehat{\mathbf{N}}^{(i)}_{j},\widehat{\mathbf{B}}^{(i)}_{j}\right\}
\end{aligned}$} \\ 
\hline
\end{tabular}
\caption{Given a discrete Ch-net we calculate the Frenet-Serret frame of the lines. The ribbons are defined as the polygons containing $\df\mathbf{U}^{(i)}_j$, $\widehat{\mathbf{B}}^{(i)}_{j}$ and $\widehat{\mathbf{B}}^{(i)}_{j+1}$.}
\label{table:discrete_frame}
\end{table}

On a surface that is not parametrized by an explicit function a Ch-net must be obtained through numerical methods, like the ones in \cite{Sageman-19, Corman-25,liu-22, Oehri-24, liu-20}. In this case the slit angles at every point, $\sigma=\sigma_{i,j}$ and $\rho=\rho_{i,j}$, are more conveniently computed by discretizing the Frenet-Serret frame. 

We adopt the simplest approach as described in the Table~\ref{table:discrete_frame}, which takes the net points ordered in lists $\mathbf{U}^{(i)}_j$ and returns the frame $\mathsf{FS}(\mathbf{U}^{(i)}_j)$, $j$ indexing the points along the $i$-th line.

Computing in the same manner the frames for the $v-$conjugate family, $\mathbf{V}^{(i)}_j$, we evaluate the slit angles at every vertex as in (\ref{eq:angles_discr}):

\begin{subequations}
    \begin{align}
\cos\sigma_{ij}&=\widehat{\mathbf{B}}^{(j)}_i(\mathbf{U})\cdot\yhwidehat{\:\:\widehat{\mathbf{N}}^{(j)}_i(\mathbf{U})\times \widehat{\mathbf{N}}^{(i)}_j(\mathbf{V})\:\:}\\
\cos\rho_{ij}&=\widehat{\mathbf{B}}^{(i)}_j(\mathbf{V})\cdot\yhwidehat{\:\:\widehat{\mathbf{N}}^{(j)}_i(\mathbf{U})\times \widehat{\mathbf{N}}^{(i)}_j(\mathbf{V})\:\:}.
    \end{align}
    \label{eq:app:discrete_prods}
\end{subequations}

\section{Bifurcations of the GMPC system}\label{sec:app-GMPC-analysis}

As noted in the text, (\ref{eq:gmpc}) is in general an overdetermined system of PDEs, although for a few particular limits of the coefficients it reduces to a single, determined equation as in (\ref{eq:sine-gordon}) or (\ref{eq:sine_gordon_flipped}). If the overdeterminacy persists it might nonetheless have solutions. In this appendix we classify all the possible solutions depending on the functional form of the inputs $\sigma(u,v),\:\rho(u,v)$. The results are summarized in the table \ref{app:table:solutions}. We find four qualitatively distinct cases: A. $\sigma,\:\rho$ vanishing, B. $\sigma,\:\rho$ different constants, C. $\sigma,\:\rho$ identical constants and D. $\sigma,\:\rho$ non-constant smooth functions. In each case the character of the system is different and its solutions depend on different arbitrary functions. The allowed form of $\widetilde{\tau}$ is likewise different.

\begin{table}[h!]
\centering
\begin{tabular}{ @{} l @{\hspace{0.4cm}} c !{\vrule width 1pt} c | c | c } 

 & Input angles & $\widetilde{\tau}=0$ & $\widetilde{\tau}=\text{{\small cnt.}}\neq 0$ &  $\mathrm{d}\widetilde{\tau}\neq 0$ \\ \Xcline{2-5}{1pt}

 {\small A} & $\sigma=\rho=0$ &  \makecell{$f_1(u),\:f_2(v)$\\{\scriptsize $K=0$}} & \makecell{$f_1(u),\:f_2(v)$\\{\scriptsize $K=-c^2$}} & - \\ \cline{2-5}

 {\small B} & {\small cnt.} $=\sigma\neq\rho=$ {\small cnt.} & $f_1(u),\:f_2(v)$ & - & - \\ \cline{2-5}

 {\small C} & $\sigma =\rho=$ {\small constant} & $f_1(u),\:f_2(v)$ & -& $c_1,\:c_2,\:c_3$ \\ \cline{2-5}

 {\small D} & {\small compatible variable} $\sigma,\:\rho$ & - & - & $c_1,\:c_2$ \\ \cline{2-5}

 {\small E} & {\small generic variable} $\sigma,\:\rho$ & - & - & - \\ \cline{2-5}
 
\end{tabular}
\caption{Classification of the nontrivial (i.e. non-constant $\gamma$) solutions of the GMPC system (\ref{eq:gmpc}). Depending on the given $\sigma,\rho$, the function $\widetilde{\tau}$ must be either constant or a varying function, and in each case the solutions depend on two one-variable initial-value functions or a few constant. Case A describes asymptotic RGSs, either planar ($\widetilde{\tau}=0$) or buckling out of plane ($\widetilde{\tau}\neq0$). If the slit angles are constant (B \& C) there is no solution unless $\widetilde{\tau}$ is identically zero or both angles equal, in the latter case all the functions depend functionally on $u\pm v$. For variable slit angles the system is genuinely overdetermined. For slit angles that are compatible with a specific configuration (D), e.g. those obtained by inverse design, $\widetilde{\tau}$ is non-vanishing and there exist solutions that depend only on two constants. For generic slit angles (E) the system admits no nontrivial solutions.}
\label{app:table:solutions}
\end{table}

\underline{\textit{Case A}}$\:-$ As shown in (\ref{eq:sine-gordon}), if $\sigma=\rho=0$ the GMPC system is reduced to a single sine-Gordon equation which has unique solutions specifying two one-variable functions, typically in the form of a Cauchy or Goursat initial conditions \cite[p.~39]{courant1962methods}. Apart from these, there is an additional constant that defines the curvature, for instance $\gamma_{uv}(0,0)$. If the grid is kept in plane $\gamma$ can still be non-constant, in which case the ribbons form two families of congruent curves.

\underline{\textit{Case B}}$\:-$ Here the situation is ostensibly different with respect to case A, however this overdetermined PDE system is consistent if and only if $\widetilde{\tau}=0$, leaving a single equation for one dependent variable as before. To show this the integrability conditions for the original system and their prolongations are written explicitly, arriving ultimately to a contradiction unless $\widetilde{\tau}$ vanishes identically. Since the derivations are largely PDE technical, a separate manuscript devoted entirely to the details of this case is in preparation \cite{castro-26-prepa}.

\underline{\textit{Case C}}$\:-$ If $\sigma=\rho=$constant we are on the one hand in the same case as before (we refer to it as case C-I) with $\widetilde{\tau}=0$ and a single determined PDE for $\gamma$.

There is, on the other hand, an additional subset of solutions (referred to as case C-II) allowed by $u\leftrightarrow v$ symmetry and not included in the previous case. We can allow for $\widetilde{\tau}\neq 0$, but it follows that $\widetilde{\tau}_u=\widetilde{\tau}_v$, and similarly we must have $\gamma_u=\gamma_v$, thus the system takes the form of two determined ODEs:

\begin{align}
    \frac{\widetilde{\tau}'}{\widetilde{\tau}^2}=A\left(\gamma\right) \:\:\:\:\:\:\:\text{and}\:\:\:\: \:\:\:\widetilde{\tau}^2=\frac{\gamma''}{\sin\gamma}+\left(\frac{ A\left( \gamma\right)\:\gamma'}{\sin\gamma}\right)^2,
    \label{eq:gmpc_one_variable}
\end{align}

being $\widetilde{\tau}=\widetilde{\tau}(u+v)$ and $\gamma=\gamma(u+v)$ single variable functions depending on three constants (say $\gamma(0), \gamma'(0),\:\widetilde{\tau}(0)$).

\begin{figure}[h!] 

\centering

  \includegraphics[width=0.6\textwidth]{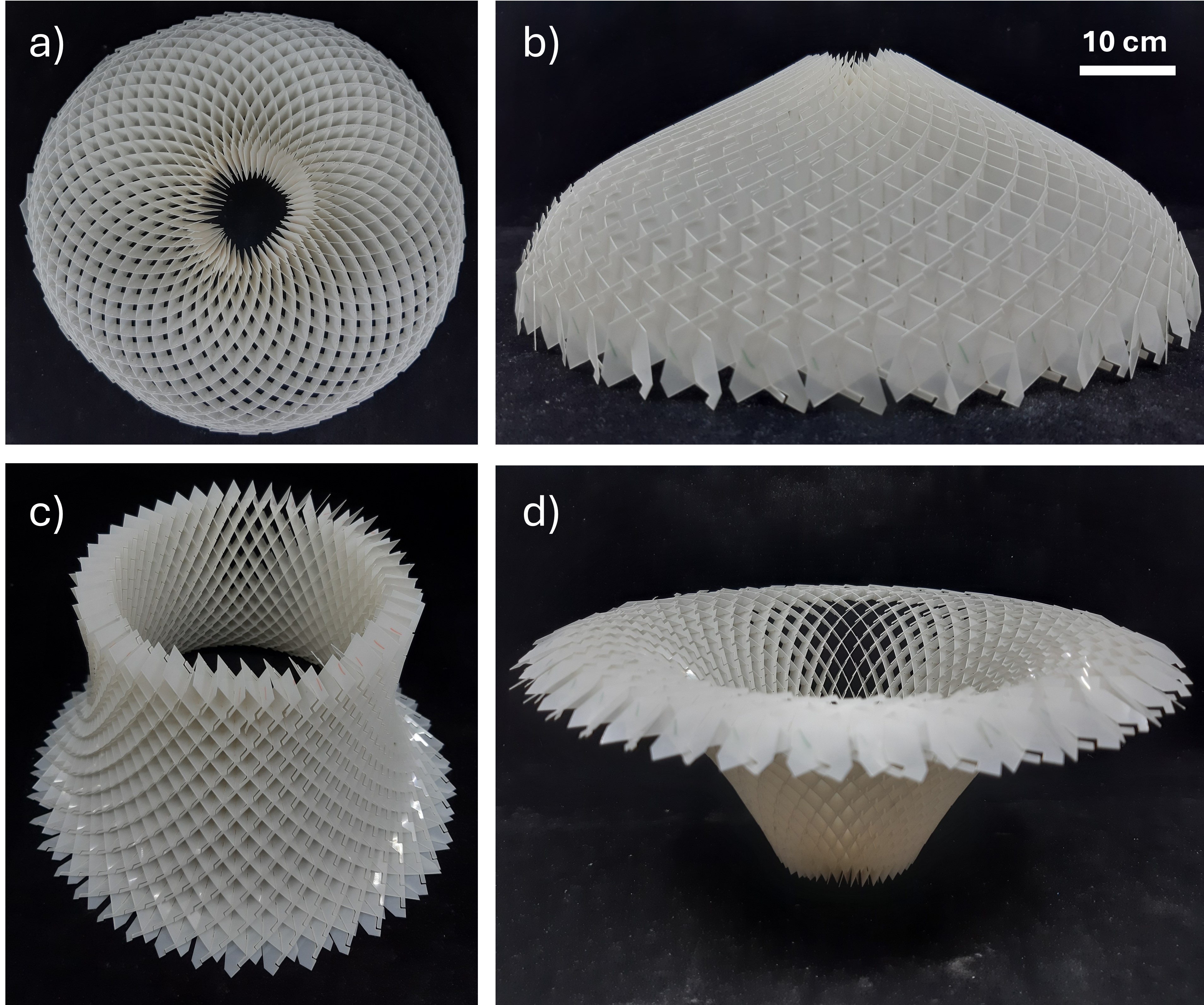}

  \caption{Axially symmetric RGS. (a) Taking identical ribbons and assembling them symmetrically we obtain the flat sheet. It can be deformed into different buckled states (b)-(d) displaying semi-rigid shapes. By construction they belong to Case C-II, where the slit angles and the functions $\gamma$ and $\widetilde{\tau}$ depend only on $u+v$, which is aligned to correspond to the vertical axis.}

  \label{fig:app:symmetric}
  
\end{figure}

An RGS described by this limiting ODE set  can be obtained assembling the ribbons symmetrically along $u-v$ lines, as shown in Fig. \ref{fig:app:symmetric}. Observe that the Gaussian curvature can change sign across the height.

\underline{\textit{Case D}}$\:-$ If $\sigma$ and $\rho$ are functions there come additional terms in the Mainardi-Codazzi equations (\ref{eq:mpc1})-(\ref{eq:mpc2}), which can be written as

\begin{subequations}
    \begin{align}
\widetilde{\tau}_u&=A\left( \gamma,\sigma(u,v),\rho(u,v)\right)\:\widetilde{\tau}^2 +\left(\frac{\partial A}{\partial \sigma}\:\sigma_v+\frac{\partial A}{\partial \rho}\:\rho_v\right)\frac{\gamma_u}{\sin\gamma}
\label{eq:c_sa_1_gen}\\
\widetilde{\tau}_v&=B\left( \gamma,\sigma(u,v),\rho(u,v)\right)\:\widetilde{\tau}^2+\left(\frac{\partial B}{\partial \sigma}\:\sigma_u+\frac{\partial B}{\partial \rho}\:\rho_u\right)\frac{\gamma_v}{\sin\gamma},
\label{eq:c_sa_2_gen}
    \end{align}
    \label{eq:app:var_mc}
\end{subequations}

while the Gauss equation remains unchanged:

\begin{align}    \widetilde{\tau}^2&=\frac{\gamma_{uv}}{\sin\gamma}+AB\:\frac{\gamma_u\gamma_v}{\sin^2\gamma}.
\tag{B.2c}
 \label{eq:app:gauss}
\end{align}

Contrary to cases B and C-I, here we must have strictly $\widetilde{\tau}\neq 0$ due to the additional terms on the right-hand side. For a variety of functional forms of $\sigma,\:\rho$ the system is found to be in involution \cite{bryant2013exterior}, and various theorems are available \cite{ivey2003cartan} to state the existence of solutions depending on two or one constants \cite{Han2003, Han2008}. Those sets of compatible $\sigma,\:\rho$ include primarily the ones constructed through inverse design.

\underline{\textit{Case E}}$\:-$ If $\sigma$ and $\rho$ are completely arbitrary functions, it will most likely be the case that (\ref{eq:c_sa_1_gen})-(\ref{eq:app:gauss}) cannot be put in involutive form, and consequently there will be no solution.

\bibliographystyle{JHEP}
 
\bibliography{biblio.bib}

\end{document}